\documentclass[prl,twocolumn,showpacs]{revtex4}
\usepackage[dvips]{graphicx}
\usepackage{epsfig}
\usepackage{amsmath,amsfonts,amssymb,bm}

\begin{document}
\title{Fluctuator model of memory dip in hopping insulators}
\author{Alexander L. Burin}
\affiliation{Department of Chemistry, Tulane University, New
Orleans, LA 70118, USA}
\date{\today}
\begin{abstract}
We show that the non-equilibrium dynamic in two-dimensional electron glasses close to metal-dielectric transition  is dramatically sensitive to electric fields confinement inside the sample, which leads to a nearly thermally activated conductance behavior and a strong non-equilibrium conductance response to the gate voltage, i. e. memory dip in a field dependence of conductance. 
Theory accounts qualitatively and quantitatively for the universal  temperature and field dependencies of memory dip. 
\end{abstract}

\maketitle


Slow relaxation in glassy materials is associated with transitions between local minima in configurational space separated by high potential barriers. At low temperature a glassy material does not approach the global energy minimum, but reaches some local minimum. There exist other local minima possessing lower energies, where the system transfers after certain expectation time. Reasonably broad distribution of transition barrier heights and lengths results in the logarithmic character of relaxation because transition rates depend on these parameters exponentially.

At low temperature ($4-20$K and below) this logarithmic relaxation is observed  in a variety of disordered materials. Particularly, a sudden application of an external electric field to amorphous dielectrics results in a fast increase of dielectric constant with its subsequent logarithmic relaxation back to equilibrium \cite{o1,ab1}. Similar behaviour of conductance is observed  after sudden  application of a gate voltage in two-dimensional Anderson insulators  including indium oxide \cite{Chorin93,Ovadyahu97,Zvi08}, granular aluminium \cite{Grenet03,Grenet07,Grenet08} and ultrathin films of Bi and Pb. \cite{Martinez97} 


A general mechanism for the non-equilibrium relaxation of conductance, earlier used to interpret experimental data for dielectric constant in glasses \cite{ab1,ab2}, can be described as following \cite{ab3,ab4}. Application of external field (gate voltage) reduces system stability in  the present local minimum and creates new pathways for its configurational transitions to lower energy states. Reduction of system energy due to configurational transitions lowers the electronic density of states (DOS)  and consequently system conductance.    

\begin{figure}[t]
\centering
\includegraphics[width=6cm]
{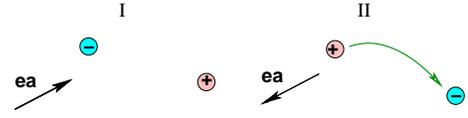}
\caption{Reduction of DOS due to the interaction of electron (blue circle with $-$) and configurational transition  (fluctuator, shown by the black arrow with a characteristic dipole moment $ea$) interaction. If the interaction of fluctuator and dipole is very large escape of electron (blue arrow), must be accompanied by the simultaneous transition of fluctuator (state I to state II). 
 \label{fig:1} }
\end{figure} 

DOS decreases due to an interaction of configurational transitions with electrons in localized states (see interaction of electron with local transition shown in Fig. \ref{fig:1}). If this interaction is sufficiently strong  then the electron needs the large energy of that interaction to leave or enter this state. At low temperature such process becomes forbidden and the only  possible excitation of electron must be accompanied by the configurational transition. Since configurational transitions occurs very rarely this electron cannot efficiently respond to the external electric field, i. e., contribute to hopping conductivity or dielectric constant. Thus equilibration of system after its disturbance, e. g. by the gate voltage, results in slow reduction of conductance by means of trapping of low energy electrons, primarily released by the gate voltage. 

\begin{figure}[t]
\centering
\includegraphics[width=6cm]
{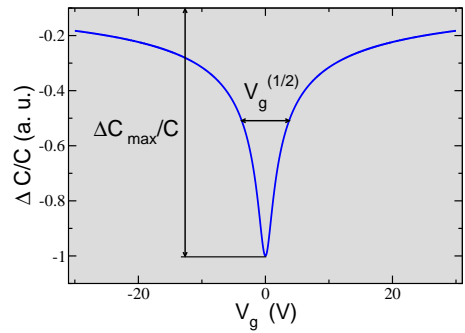}
\caption{Typical shape of memory dip
 \label{fig:2} }
\end{figure}

Another interesting observation related to glassy properties is the memory dip seen in conductance after a fast sweep of the gate voltage around its equilibrium value, $V_{g}=0$ (see $V$-shaped conductance dependence on that voltage in Fig. \ref{fig:2}). The greater the voltage the stronger the system departs from equilibrium increasing the conductance, according to the previously described scenario. Memory dip can be characterized by its half-width $V^{(1/2)}_{g}$ and depth $\delta C_{max}/C$. 

The temperature dependence of conductance, $C$, memory dip parameters,  $V^{(1/2)}_{g}$, and $\delta C_{max}/C$, and logarithmic relaxation rate of conductance, $r_{ln}=d\ln(C)/d\ln(t)$, contain significant information about the nature of the glassy state. It turns out that within the temperature range of interest, $4$K$\leq T\leq 25$K, the conductance of  most of  samples shows nearly thermally activated behavior, $C \propto e^{-(T_{0}/T)^{n}}$, $0.8\leq n \leq 1$ \cite{Grenet08,Zvi08,Martinez97}. This behaviour can be interpreted as a consequence of field confinement in $2$ dimensions ($2D$)  \cite{Shkl2Dhopp,Old2Dhopp}, rather than the standard variable range hopping mechanism, $1/4\leq n \leq 1/2$, considered in earlier work \cite{ab3,ab4,abnoise,cg1,cg2,cg3,cgSG,cg5}. In this letter we suggest theoretical model of electronic glass involving the $2D$ field confinement, which is capable to interpret several observed behaviors at intermediate temperature, $T\geq 4K$, including (A) linear temperature dependence of a memory dip half-width and its universality with respect to a sample conductance \cite{Grenet08} (cf. \cite{Zvi08}); (B) nearly $T^{-2}$ dependence of $\Delta C/C$ varying for samples with different conductances \cite{Grenet08}; and (C) nearly $T^{-3/2}$ temperature dependence of logarithmic relaxation rate \cite{Martinez97}. Absolute values of these parameters agree with the common sense expectations.

If the dielectric constant of the film of interest, $\kappa_{in}$ exceeds that of the environment, $\kappa_{in}\gg \kappa_{ex1}=(\kappa_{1}+\kappa_{2})/2$ ($\kappa_{1}$, $\kappa_{2}$ are dielectric constants of surrounding materials),  the interaction of electrons differs from the standard Coulomb interaction, $e^2/(\kappa_{in}r)$,  because of the field confinement within the layer \cite{Old2Dhopp,Shkl2Dhopp} (see also Ref. \cite{Zvi86}, where the field confinement was considered for Ti oxide based materials). In that case the interaction of two electrons at short distances $r<d/2$ can be expressed with the logarithmic accuracy as 
\begin{equation}
U_{C}(r)=\frac{e^2}{\kappa_{in}r}+2\Delta_{*},  ~ \Delta_{*}\approx\frac{e^2}{\kappa_{in}d}\ln\left(\frac{\kappa_{in}}{\kappa_{1}}\right). 
\label{eq:CoulConf1}
\end{equation}
This interaction results in the hard Coulomb gap $\Delta_{*}$ in a DOS leading to the conductance behaviour \cite{Old2Dhopp,Shkl2Dhopp}
\begin{equation}
\sigma= \sigma_{0}e^{-\frac{\Delta_{*}}{k_{B}T}-\sqrt{\frac{E_{0}}{k_{B}T}}},   ~ E_{0}=2.8\frac{e^{2}}{\kappa_{in}a},  
\label{eq:CoulConf}
\end{equation}  
where $a$ stands for the electron localization length. 

This dependence is similar to the one reported in Ref. \cite{Grenet08} for various granular aluminium samples in the temperature domain of interest $4$K $<T<20$K and the most significant temperature dependence in exponent is associated with the Arrhenius term in the exponent  $\Delta_{*}/(k_{B}T)$. Consequently, one can expect that the non-equilibrium conductance behavior is also associated with this term. Then  the relaxation and memory dip are determined by the dielectric constant time and field dependencies as 
\begin{equation}
\frac{\delta \sigma(V_{g}, t)}{\sigma} \approx \frac{\Delta_{*}}{k_{B}T}\frac{\delta \kappa_{in}(V_{g}, t)}{\kappa_{in}}. 
\label{eq:NoneqCond}
\end{equation} 
It is quite natural to expect that  the dielectric constant of Anderson insulators behaves similarly to that in amorphous solids \cite{o1,ab1,ab2} yielding experimentally observed conductance behavior.    


To characterize the non-equilibrium dynamics one has to describe configurational transitions.  In dielectric glasses relaxation and ageing phenomena are associated with the two-level systems (TLS)\cite{AHVP} formed by tunneling transitions of atoms or groups of atoms between two close energy minima separated by potential barriers. TLS possess a universal statistics with respect to their energies, $E$, and relaxation times, $\tau$, 
\begin{equation}\label{eq:TLS}
P(E, \tau) =\frac{P_{*}}{\tau}.
\end{equation}
and a small dipole moments $\mu_{str} \approx 1-3$D. 

One can expect that in Anderson insulators electrons themselves can form the electron glass state \cite{cg2,cgSG,cg5,EG1,EG3,Clare}.  Assuming that electronic glass is formed due to Coulomb interaction one can estimate the glass transition temperature, $T_{g}$ is given by $k_{B}T_{g}=E_{0}$ (see Eq. (\ref{eq:CoulConf}), cf. Ref. \cite{cgSG}). In this state the slow dynamics is associated with collective electronic transitions between close energy minima. The logarithmic relaxation can be due to exponential sensitivity of electron transition rate to the hopping length and the number of participating electrons \cite{ab4,abnoise}.  Transitions observable within the time domain of the typical experiment, i. e., from several seconds to several weeks, cannot involve more than   $5-10$ electrons \cite{abnoise} located close to each other. Then we can still use a TLS like model Eq. (\ref{eq:TLS}) to characterize electronic transitions as local fluctuators \cite{abnoise} possessing density, $P_{}$ and dipole moments $\mu_{*}\sim ea$; the localization length $a$ is the only length parameter of the problem. It is convenient to characterize fuctuators by the dimensionless product $P_{*}\mu_{*}^2/\kappa_{in} \approx \chi_{*}$, so one has 
\begin{equation}\label{eq:TLSe}
\mu_{*} \approx ea, ~ P_{*} \approx \frac{\chi_{*}\kappa_{in}}{e^2 a^2}.
\end{equation}
In amorphous solids $\chi_{*}$ is a universal parameter of order of $10^{-3}-10^{-4}$, which can possess weak (logarithmic) temperature dependence. As we will see below, the similar assumption leads to the right estimate of the memory dip depth in conductance. Although the mechanism of electron glass formation is beyond the scope of the present work we believe that the long range $3-D$ interaction is important there (cf. \cite{ab1}). Therefore the formation of glassy state can be much less efficient in the case of small concentration of carriers (less than one per cubic thickness of material) in agreement with experimental observations \cite{Zvi08}. Also if localization length exceeds the thickness the effective interaction is two-dimensional, which can also suppress the glass formation \cite{abnoise}.

Consider the effect of fluctuators on the dielectric constant determining the non-equilibrium behavior Eq. (\ref{eq:NoneqCond}). The dielectric constant  is associated with the polarization of electron-hole pairs of various lengths $r$, smaller than the hopping length. Similarly to Ref. \cite{ab2} the part of dielectric constant most sensitive to its interaction with fluctuators is associated with low-energy electron-hole pairs  separated by intermediate distance, $r$, such that  $a<r<d/2$. It is harder to disturb shorter dipoles due to their large tunneling splitting, $\Delta_{0} \geq k_{B}T_{0} \gg k_{B}T$, while longer pairs do not make important contribution due to the ``hard'' Coulomb gap at corresponding energies caused by field confinement in $2D$. Let those dipoles be characterized by the density function $F(\Delta, r)$ depending on their energy, $\Delta$ and size $r$. This size determines the tunneling amplitude of electron between two localized states, $\Delta_{0}\approx E_{0} e^{-r/a}$.  Then one can approximate the contribution of interest to the dielectric constant as (c. f. \cite{ESreview,GalperinAbs})
\begin{eqnarray}
\Delta \kappa_{in} \approx \frac{2\pi e^2}{3}\int_{a}^{d/2} r^2 dr 
 \int_{-\infty}^{\infty}d\Delta \frac{\Delta_{0}^2 F(\Delta, r)}{E^{2}} 
\nonumber\\
\times \left(  \frac{1}{E}\tanh\left(\frac{E}{2k_{B}T}\right)+  \frac{\Delta^2}{2k_{B}T\Delta_{0}^{2}}\cosh^{-2}\left(\frac{E}{2k_{B}T}\right) \right);
 \nonumber\\
 E=\sqrt{\Delta^2+\Delta_{0}^2}. 
\label{eq:DielConst}
\end{eqnarray}  
Two contributions in the right hand side of Eq. (\ref{eq:DielConst}) are identical to resonant and relaxational contributions to the dielectric constant in amorphous solids \cite{ab1,ab2} and the latter contribution is more significant because it is determined by lowest energies of pairs.

In the absence of fluctuators the density function of sufficiently large dipoles, $a<r<d$ is given by \cite{ESreview,GalperinAbs}
\begin{equation}
F(\Delta, r)\approx \frac{3}{10\pi^2}\left(\frac{\kappa_{in}}{e^2}\right)^6\left(\frac{e^2}{\kappa_{in}r}+\Delta\right)^{5}. 
\label{eq:DensDip}
\end{equation}
The interaction of dipoles with fluctuators reduces their density of states \cite{ab2,ab4}. Only interaction exceeding the thermal energy is important. We assume, that the distance between the fluctuator and closest charge of the dipole is smaller than the size of the dipole $r$, which is necessary to make the interaction larger than the thermal energy. This is true for $r\sim d/2$ and $T>4$K \cite{Grenet07}. Then the correction to the dipole density can be expressed as \cite{ab3,ab4} (remember, $ E=\sqrt{\Delta^2+\Delta_{0}^2}$) 
\begin{equation}
  \label{eq:004}
\frac{\delta F(\Delta, r)}{F(\Delta, r)} \approx  -\frac{32\pi}{15}  P_{*}E_{0}a^3
  \left[\frac{E_{0}}{ E }\right]^{1/2}
  \ln \left(\frac{t}{\tau_{min}}\right).
\end{equation}
The time dependence is due to a logarithmically uniform distribution of fluctuator relaxation time (Eq. (\ref{eq:TLS})), $t$ stands for the observation time and $\tau_m$ is some characteristic minimum time, determined by the rate of a gate voltage application. 

Substituting Eq. (\ref{eq:004}) into Eq. (\ref{eq:DielConst}), performing the straightforward integrations and substituting the final result into Eq. (\ref{eq:NoneqCond}) we obtain the following expression for the fluctuator correction to the conductance 
\begin{eqnarray}
\frac{\delta C}{C} \approx - \frac{32\sqrt{2}\chi_{*}}{75}  \frac{\Delta_{*}}{k_{B}T}\sqrt{\frac{E_{0}}{k_{B}T}}\ln\left(\frac{d}{2a}\right)\ln \left(\frac{t}{\tau_{}}\right).
\label{eq:CorrDielConst}
\end{eqnarray}  
This equation describes the ageing effect as well as the conductance time dependence the application of a large gate voltage bringing all relevant fluctuators out of equilibrium. The temperature dependence of the conductance logarithmic relaxation rate  $r=d\ln(C)/d\ln(t)$ is close to the power law $r\propto T^{-3/2}$ that is consistent with the observations of Ref. \cite{Martinez97} in ultrathin films of Bi and Pb.

Consider the shape of the memory dip (Fig. \ref{fig:1}). The energy change of fluctuator with the dipole moment $\mu$ associated with the application of a gate voltage $V_{g}$ is given by $\delta E = F_{g}\mu \cos(\theta)\approx V_{g}\kappa_{1}\mu \cos(\theta)/(\kappa_{in}d_{tot})$, where $\theta$ is the angle between directions of the dipole moment and external electric field $F_{g}$, $d_{tot}$ is the distance between sample and gate electrode and $k_{1}$ is the dielectric constant of the insulating layer separating the gate electrode from the sample (see Refs. \cite{Grenet08,Zvi08} for details). Only fluctuators with sufficiently small energy $E$, $0<E<\delta E$, will be really disturbed from equilibrium. This condition sets the upper constraint $\delta E$ for integration over energy in Eq. (\ref{eq:DielConst}). Also the time dependent logarithm $\ln (t_{max}/t_{min})$ in Eq. (\ref{eq:DielConst}) should be modified. It is determined by the contribution of fluctuators removed from equilibrium by the gate voltage, which possess the relaxation time smaller than  the maximum time $t_{max}$, determined by the time the sample was kept at fixed temperature after cooling and larger than the minimum time, $t_{min}$, determined by the gate voltage sweep rate, $r_{s}$, as $t_{min} \sim k_{B}T/(er_{s})$. 
Then the memory dip can be described by the equation   
\begin{eqnarray} 
\frac{\delta C(V_{g})}{C} \approx -\frac{32\sqrt{2}\chi_{*}}{75} \frac{\Delta_{*}}{k_{B}T}\ln\left(\frac{d}{2a}\right)
\nonumber\\
\times\ln \left(\frac{t_{max}r_{s}e\kappa_{1}}{k_{B}Td_{tot}\kappa_{in}}\right)\left[\sqrt{\frac{E_{0}}{k_{B}T}}-2\sqrt{\frac{E_{0}d_{tot}\kappa_{in}}{V_{g}ea\kappa_{1}}}\right]. 
\label{eq:MemDip}
\end{eqnarray}
Using this result one can estimate the half width of the memory dip as 
\begin{equation}
V^{(1/2)}_{g}\approx 16 \frac{k_{B}T}{e} \frac{\kappa_{in}d_{tot}}{\kappa_{1}a}. 
\label{eq:half_width}
\end{equation}

This result agrees with the experimentally observed linear temperature dependence of the memory dip width at temperatures $T\geq 4$K \cite{Grenet08,Zvi08}. At lower temperatures the dependence is getting stronger. In our opinion this is because the estimate for gate voltage induced electric field, $F_{g} \approx V_{g}^{(1/2)}\kappa_{1}/(\kappa_{in}d_{tot})$, is no longer valid. One can use that macroscopic expression  if the density of electrons injected due to the application of a gate voltage exceeds one electron per the cubic sample thickness. This is satisfied for granular Aluminium samples down to $4$K \cite{Grenet03}, but fails at lower temperatures, where this field is defined by the injected electron, closest to the given fluctuator. One can show that at low temperatures a width of memory dip is proportional to squared temperature, $V_{g}^{(1/2)}\sim \frac{(k_{B}T)^2}{e}\frac{d^{2}d_{1}\kappa_{in}^2}{e^2 \kappa_{1}a^2}$. 

The quantitative universality of the memory dip halfwidth  discovered in \cite{Grenet08} for samples with conductances different by orders of magnitude requires understanding. It is important  that in the vicinity of a metal-insulator transition the ratio of two diverging parameters, $\epsilon_{in}/a$, is expected to be approximately constant on the dielectric side both in accordance with the theoretical analysis and experimental data \cite{BorisScaling}. One can naively expect that at the scale of localization length  the characteristic kinetics and Coulomb energies should be close to each other, $E_{0}=e^{2}/(\kappa_{in}a)\approx \hbar^{2}/(ma^{2})$ which, results in the universal ratio $a/\epsilon_{in} \sim \hbar^2/(me^2) \approx 0.5\AA $. This is simply the Bohr's radius, $a_{B}$ ($m$ is the effective mass of electron, which is taken to be equal to the bare electronic mass). 

Under these assumptions Eq. (\ref{eq:half_width}) results in the universal dependence of the memory dip halfwidth on the gate voltage $V^{(1/2)}_{g} = \eta T$ with the proportionality coefficient coefficient $\eta$ depending only on the thickness $d_{tot}$ and dielectric constant $\kappa_{1}$ of the layer separating the sample and the gate electrode. For granular Aluminium sample one has $\kappa_{1}\approx 9$, $d_{tot}=1000\AA$ which yields $\eta=0.29 $V K$^{-1}$ in excellent agreement with the experimental result $\eta \approx 0.25$V K$^{-1}$ \cite{Grenet08}. 

The same approach does not work so good for the half-width of the memory dip reported in Ref. \cite{Zvi08} for In$_{2}$O$_{3-x}$ sample having a relatively small resistance of $200$k$\Omega$. We believe that this is because the localization length in this almost conducting sample exceeds the thickness of the sample $d_{InO}\sim 30 \AA$ and therefore one should replace the localization length $a$ in Eq. (\ref{eq:half_width}) with the sample thickness. Assuming that the localization length exceeds the sample thickness by an order of magnitude one can make the experimental data consistent with the theory. Other data reported in Ref. \cite{Zvi08} correspond to very low temperatures. However, in granular Aluminium  one can estimate (fitting activation energy $20-40$K with  Eq. (\ref{eq:CoulConf1}) and using the relationship $a\approx 0.5 \kappa_{in} (\AA)$)  that $15\AA < a < 50\AA$ in all samples of thickness $200 \AA$ so our estimate is justified.  

The temperature dependence of the depth of the memory dip in Eq. (\ref{eq:MemDip}) in the limit of $V_{g}\rightarrow \infty$ is determined by power law temperature dependence $T^{-3/2}$, possible weak temperature dependence of the dimensionless parameter $\chi_{*}$  and the logarithmic factor $\ln \left(\frac{t_{max}r_{s}k_{ex}}{k_{B}Td_{tot}\kappa_{in}}\right)$. Integration of these three dependences could be responsible for the $T^{-2}$ behavior of the memory dip depth  reported in Ref. \cite{Grenet07}. 

According to the experiment \cite{Grenet07} the absolute value of the relative correction reaches few percents at $T \sim 4K$ for less conducting sample, which is consistent with Eq. (\ref{eq:MemDip}) if one set $\chi_{*} \sim 5\cdot 10^{-4}$ in agreement with the typical values of that parameter typical value in amorphous solids \cite{ab1}.

Thus we suggested the model of electronic glass. In this model slow dynamics can be characterized by fluctuators similar to TLS in amorphous solids.  Our model is consistent with existing experimental data both qualitatively and quantitatively. Theoretical predictions, Eqs. (\ref{eq:MemDip}), ((\ref{eq:half_width}) can be tested varying system parameters including electron localization length, sample thickness and temperature.

\acknowledgements
This work is partially supported by the Tulane Research and Enhancement Fund. Author is grateful to Zvi Ovadyahu and Thierry Grenet for useful discussions.

\end{document}